\documentclass[review]{elsarticle}

\usepackage{amsmath,lineno,hyperref}
\modulolinenumbers[5]

\journal{Journal of \LaTeX\ Templates}

\usepackage{color}
\usepackage[normalem]{ulem}

\renewcommand\sout{\bgroup \color{red} \ULdepth=-.5ex \ULset}
\newcommand{\bs}{\boldsymbol}

\begin{document}

\begin{frontmatter}

\title{Mesoscopic dynamics of fermionic cold atoms\\
  --- Quantitative analysis of transport coefficients and relaxation times ---}


\author[mymainaddress]{Yuta Kikuchi}
\ead{kikuchi@ruby.scphys.kyoto-u.ac.jp}

\author[mysecondaryaddress]{Kyosuke Tsumura}

\author[mymainaddress]{Teiji Kunihiro}

\address[mymainaddress]{Department of Physics, Faculty of Science, Kyoto University,
Kyoto 606-8502, Japan.}
\address[mysecondaryaddress]{Analysis Technology Center,
  Research \& Development Management Headquarters,
  Fujifilm Corporation,
  Kanagawa 250-0193, Japan.}

\begin{abstract}
We give a quantitative analysis of 
the dynamical properties of fermionic cold atomic gases in normal phase, such as
 the shear viscosity, heat conductivity, and viscous relaxation times, 
using the novel microscopic expressions derived by the renormalization group (RG) method,
where the Boltzmann equation is faithfully solved
to extract the hydrodynamics without recourse to any ansatz.
In particular, we examine 
the quantum statistical effects, temperature dependence, and scattering-length 
dependence of the transport coefficients and the viscous relaxation times. 
The numerical calculation shows that the relation $\tau_\pi=\eta/P$, 
which is derived in the relaxation-time approximation (RTA) and is used in most of the literature, 
turns out to be satisfied quite well, while the similar relation for
the viscous relaxation time $\tau_J$ of the heat conductivity is satisfied only approximately with a considerable error.
\end{abstract}


\end{frontmatter}


\section{Introduction}

\label{sec:sec1}
The viscous hydrodynamic equation provides us with a unified way to describe the near-equilibrium systems
in terms of macroscopic quantities such as the pressure,
 particle number density, and fluid velocity.
 The detailed microscopic properties of the system are renormalized into transport coefficients
 such as the shear viscosity, heat conductivity and so on. 
Therefore, the elaborate investigation of the transport coefficients
is one of the most important tasks to reveal the microscopic properties of the fluid. 
For instance, the fluid with a tiny shear viscosity is realized in the experiment of 
ultracold Fermi gases at unitarity
 \cite{Ohara:2002observation,Kinast:2004,Bartenstein:2004,Schafer:2007,Cao:2010wa,Elliott:2014}: 
The value of its shear viscosity is close to a quantum bound that is theoretically 
proposed \cite{Policastro:2001yc,Kovtun:2004de}, 
implying the realization of the strongly correlated systems at the  unitarity
\cite{Gelman:2005,Bruun:2007etas,Rupak2007,Enss:2011,Hao:2011,Enss:2012}.

Since the naive hydrodynamic description, however, breaks down 
when the mean free path is not small enough compared with the macroscopic length scale,
we should incorporate more microscopic properties such as
the relaxation process of dissipative currents, which is characterized by viscous relaxation 
times \cite{Massignan:2005,Bruun:2005,Bruun:2007etas,Bruun:2007,Braby:2011,Chao:2012}. 
Dynamical equation taking into account them may be called 
{\it mesoscopic} \cite{jou1996extended,dedeurwaerdere1996foundations} 
since it describes the dynamics in an intermediate scale
between those described by the Navier-Stokes and the Boltzmann equations.
The second-order hydrodynamic equation describes the mesoscopic dynamics including the relaxation 
of the dissipative currents, in addition to the ordinary hydrodynamic behavior described 
by the Navier-Stokes equation.
In particular, the quantitative estimates of the viscous relaxation times are crucial 
to predict beyond-Navier-Stokes behavior of fluids in the mesoscopic regime, and
such a behavior should be realized in the dilute cold atomic gases, for instance.
Indeed the importance of the second-order hydrodynamics
 has been recognized and many attempts have been done to derive it
\cite{levermore1996moment,karlin1998dynamic,struchtrup2003regularization,gorban2004invariant,torrilhon2009special,torrilhon2010hyperbolic},
and among them the renormalization group (RG) method seems 
to be a promising method to derive the second-order hydrodynamics 
as well as the transport coefficients including the viscous relaxation times.
In fact, it has been applied to derive the second-order hydrodynamic equations 
from the Boltzmann equation for both relativistic and non-relativistic systems
 \cite{Tsumura:2012gq,Tsumura:2012kp,Tsumura:2013uma,Tsumura:2015fxa}, 
and desirable properties have been already proved for the resultant equation such as causality, 
stability, positivity of the entropy production rate, 
and the Onsager's reciprocal relation without imposing any assumption a priori
 \cite{Tsumura:2015fxa,Kikuchi:2015swa}. 
Moreover, it is worth emphasizing that the microscopic expressions obtained for the transport coefficients 
such as the shear viscosity, heat conductivity and  so on coincide with those derived in the
celebrated Chapman-Enskog method, while the novel  microscopic expressions
of the viscous relaxation times written in terms of the relaxation functions
allow physically natural interpretations as the relaxation times.

The aim of this article is to give basic analyses of nonequilibrium properties 
of cold  atomic gasses in a quantitative way based 
on the kinetic theory  
\cite{Tsumura:2012gq,Tsumura:2012kp,Tsumura:2013uma,Tsumura:2015fxa};
a focus is put on a quantitative and comprehensive analysis of the viscous relaxation times
based on the novel  microscopic formulas of them \cite{Tsumura:2012gq,Tsumura:2012kp,Tsumura:2013uma,Tsumura:2015fxa}.
Such an analysis is of fundamental importance in this field because
the quantitative extraction of the relaxation times 
is indispensable to elucidate the  dynamics of cold atomic gases in a precise way. 
In focusing nonequilibrium properties, we shall not take care
of the pairing correlations including the superfluid transition nor medium effects, 
which are extensively discussed in 
Refs.~\cite{Gelman:2005,Bruun:2007etas,Rupak2007,Enss:2011,Hao:2011,Enss:2012,Wlazlowski:2012jb,Wlazlowski:2013owa,Wlazlowski:2015yga}.
Instead, we perform a quantitative test of the shear viscosity, heat conductivity, and viscous relaxation times 
of the stress tensor and heat flow using the microscopic expressions obtained in RG method applied Boltzmann equation, 
and examine the temperature and scattering-length dependence 
and quantum statistical effects: The evaluation of these quantities are converted to 
solving the corresponding linear integral equations, which we solve numerically without 
recourse to any approximation.
On the basis of the numerical results, we also examine how well the relation $\tau_\pi=\eta/P$ and 
its analog for the heat conductivity \cite{Bruun:2007,Braby:2011,Chao:2012} are satisfied, 
where $\tau_\pi$ is the viscous relaxation time of the stress tensor, $\eta$ is the shear viscosity, and $P$ is the pressure.
These relations are obtained from the Boltzmann equation with use of the relaxation-time approximation (RTA), 
which has been widely applied to a lot of studies of the kinetic theory.
Quantitative reliability of the RTA is unclear and has  been hardly checked:
We are only aware of \cite{Schafer:2014} in which the validity of the RTA is analytically examined up 
to some approximations.
The present analyses confirm that the quantum statistical effect significantly contribute to the transport 
coefficients at low temperature and near unitarity.
Furthermore, we find that the relation $\tau_\pi=\eta/P$
is well satisfied quantitatively, while the analogous relation for
the viscous relaxation time of the heat conductivity $\tau_J$ is not satisfied well.



\section{Hydrodynamic equation derived by RG method}
\label{sec:sec2}
The RG method is a general framework 
to identify slow variables and extract their dynamics 
from original theories \cite{Chen:1994zza,Kunihiro:1995zt,Kunihiro:1996rs,Ei:1999pk}.
In the RG method, the hydrodynamic equation is derived by faithfully solving the Boltzmann equation given by
\begin{align}
 \label{eq:BE1}
 \left(\frac{\partial}{\partial t} + \bs{v}\cdot \bs{\nabla}\right)
 f_p(t,\bs{x})=C[f]_p(t,\bs{x}),
\end{align}
with $\bs{v}\equiv \bs{p}/m$ and the collision integral given by
\begin{align}
 \label{eq:collision}
 C[f]_p(t,\boldsymbol{x})
 &=\frac{1}{2}\int_{p_1}\int_{p_2}\int_{p_3}\mathcal{W}(p,p_1|p_2,p_3)
 \nonumber\\
 &\times\big(\bar{f}_p\bar{f}_{p_1}f_{p_2}f_{p_3}-f_p f_{p_1}\bar{f}_{p_2}\bar{f}_{p_3}\big).
\end{align}
Here, we have introduced the notations $\int_p\equiv \int\mathrm{d}^3p/(2\pi)^3$
 and $\bar{f}\equiv 1+af$. $a$ represents quantum statistics, i.e., $a=-1(+1)$ 
for fermion (boson) and $a=0$ for the classical Boltzmann gas.
$\mathcal{W}$ is a transition matrix given by
\begin{align}
 \label{eq:transition}
 \mathcal{W} &= |\mathcal{M}|^2 (2\pi)^4\delta(E+E_1-E_2-E_3)
 \nonumber\\
 &\times\delta^3(\bs{p}+\bs{p}_1-\bs{p}_2-\bs{p}_3),
\end{align}
with $E=|\bs{p}|^2/(2m)$ and the scattering amplitude $\mathcal{M}$.
For later convenience, we define the linearized collision operator,
\begin{align}
 L_{pq}\equiv
 (f^{\mathrm{eq}}_p\bar{f}^{\mathrm{eq}}_p)^{-1}
 \left.\frac{\delta}{\delta f_q}C[f]_p(t)\right|_{f=f^{\mathrm{eq}}}
 f^{\mathrm{eq}}_q\bar{f}^{\mathrm{eq}}_q.
\end{align}

The resultant hydrodynamic equations take the familiar forms
\begin{align}
 \label{eq:cont-eq-n3}
 \frac{\mathrm{D}n}{\mathrm{D}t} &= -n\boldsymbol{\nabla}\cdot\boldsymbol{u},
 \\
 \label{eq:cont-eq-u3}
 mn\frac{\mathrm{D}u^i}{\mathrm{D}t} &= -\nabla^i P+ \nabla^j \pi^{ij},
 \\
 \label{eq:cont-eq-s3}
 Tn\frac{\mathrm{D}s}{\mathrm{D}t} &= \boldsymbol{\nabla}\cdot\boldsymbol{J} + \sigma^{jk}\pi^{jk},
\end{align}
where $n$, $\bs{u}$, $m$, $P$, and $s$ are 
the particle number density, fluid velocity, mass density, pressure, and entropy density, 
respectively. We have introduced the Lagrange derivative defined by $\mathrm{D}/\mathrm{D}t\equiv\partial/\partial t+\bs{u}\cdot\bs{\nabla}$ and the shear tensor defined by $\sigma^{ij}\equiv \frac{1}{2}\nabla^i u^j+\frac{1}{2}\nabla^j u^i-\frac{1}{3}\delta^{ij}\nabla^k u^k$.
As worked out in Refs.~\cite{Tsumura:2013uma} and \cite{kikuchi:2015PTEP} with classical and quantum statistics, respectively, the relaxation of the stress tensor $\pi^{ij}$ and heat flow $J^i$ is described by the following relaxation equation,
\begin{align}
 \label{eq:relax1}
  \pi^{ij}
  &= 2\eta\sigma^{ij}
  - \tau_\pi \frac{\mathrm{D}}{\mathrm{D}t}\pi^{ij}
  - \ell_{\pi J}\nabla^{\langle i} J^{j\rangle} 
  \nonumber\\
  &+ \kappa_{\pi\pi}^{(1)}\pi^{ij}\boldsymbol{\nabla}\cdot\boldsymbol{u}
 + \kappa_{\pi\pi}^{(2)}\pi^{k\langle i}\sigma^{j\rangle k}
 - 2\tau_\pi \pi^{k\langle i}\omega^{j\rangle k}
 \nonumber\\
 &+ \kappa_{\pi J}^{(1)}J^{\langle i}\nabla^{j\rangle}n
 + \kappa_{\pi J}^{(2)}J^{\langle i}\nabla^{j\rangle}P
  \nonumber\\
  &+ b_{\pi\pi\pi} \pi^{k\langle i} \pi^{j\rangle k}
  + b_{\pi JJ} J^{\langle i} J^{j\rangle},
\\[5pt]
  \label{eq:relax2}
 J^i
  &=\lambda \nabla^i T
  -\tau_J \frac{\mathrm{D}}{\mathrm{D}t}J^i
  -\ell_{J\pi}\nabla^j \pi^{ij}
  \nonumber\\
  &+ \kappa_{J\pi}^{(1)}\pi^{ij}\nabla^j n
 + \kappa_{J\pi}^{(2)}\pi^{ij}\nabla^j P
 \nonumber\\
 &+ \kappa_{JJ}^{(1)}J^i\boldsymbol{\nabla}\cdot\boldsymbol{u}
 + \kappa_{JJ}^{(2)}J^j\sigma^{ij}
 + \tau_J J^j\omega^{ij}
  \nonumber\\
  &+ b_{JJ\pi}J^j \pi^{ij}.
\end{align}
Here for an arbitrary tensor $A$, $A^{\langle ij\rangle}\equiv \Delta^{ijkl}A^{ij}$ 
with the definition of a symmetric traceless tensor 
$\Delta^{ijkl}\equiv \frac{1}{2}\delta^{ik}\delta^{jl}+
\frac{1}{2}\delta^{il}\delta^{jk}-\frac{1}{3}\delta^{ij}\delta^{kl}$.
The microscopic expressions of the shear viscosity and heat conductivity are given in terms of the linearized collision operator $L$ by
\begin{align}
 \label{eq:TC}
 \eta \equiv -\frac{1}{10T}\langle\hat{\pi}^{ij},L^{-1}\hat{\pi}^{ij}\rangle,
 \ \ \ 
 \lambda \equiv -\frac{1}{3T^2}\langle\hat{J}^i,L^{-1}\hat{J}^i\rangle,
\end{align}
while the viscous relaxation times are given by
\begin{align}
 \label{eq:RT}
 \tau_{\pi} \equiv \frac{1}{10T\eta}\langle\hat{\pi}^{ij},L^{-2}\hat{\pi}^{ij}\rangle,
 \ \ \
 \tau_{J} \equiv \frac{1}{3T^2\lambda}\langle\hat{J}^{i},L^{-2}\hat{J}^{i}\rangle.
\end{align}
with the definition of the inner product given by
\begin{align}
 \left<\psi,\chi\right>\equiv 
 \int_p f^{\mathrm{eq}}_p\bar{f}^{\mathrm{eq}}_p\psi_p\chi_p.
\end{align}
with the definition of the microscopic expressions of the stress tensor and heat flow,
\begin{align}
 \hat{\pi}^{ij}_p &\equiv \delta v^{\langle i}\delta p^{j\rangle},
 \ \ \ 
 \hat{J}^i_p \equiv \left(\frac{|\bs{\delta p}|^2}{2m}-h\right)\delta v^i,
\end{align}
respectively.
The relative momentum $\bs{\delta p}$ and relative velocity $\bs{\delta v}$ against the fluid velocity $\bs{u}$ are defined by $\bs{\delta p} \equiv m\bs{\delta v} \equiv m(\bs{v}-\bs{u})$ and $h$ is the enthalpy density.
The meanings and explicit expressions of the other coefficients 
such as $\ell_{\pi J}$, $\kappa_{\pi\pi}^{(1)}$, and $b_{\pi\pi\pi}$ are given in Ref.~\cite{kikuchi:2015PTEP} 
with an external force taken into account 
(see also \cite{Tsumura:2012gq,Tsumura:2012kp,Tsumura:2015fxa,Kikuchi:2015swa}, for applications to relativistic systems). 
The hydrodynamic equations \eqref{eq:cont-eq-n3}--\eqref{eq:relax2},
 the transport coefficients and viscous relaxation times \eqref{eq:TC} and \eqref{eq:RT} 
have been already derived in Ref.~\cite{Tsumura:2013uma} for classical Boltzmann gases.
Defining ``time-evolved" vectors $\{\hat{\pi}^{ij}_p(s),\hat{J}^{ij}_p(s)\} \equiv \{[\mathrm{e}^{sL}\hat{\pi}^{ij}]_p,[\mathrm{e}^{sL}\hat{J}^{ij}]_p\}$, we may convert Eqs.~\eqref{eq:TC} and \eqref{eq:RT} into the following forms
\begin{align}
 \label{eq:TC1-2}
 \eta &\equiv \frac{1}{10T}\int_0^\infty\mathrm{d}s\langle\hat{\pi}^{ij}(0),\hat{\pi}^{ij}(s)\rangle,
 \\
 \label{eq:TC2-2}
 \lambda &\equiv \frac{1}{3T^2}\int_0^\infty\mathrm{d}s\langle\hat{J}^i(0),\hat{J}^i(s)\rangle,
 \\
  \label{eq:RT1-2}
 \tau_{\pi} &\equiv \frac{\int_0^\infty\mathrm{d}s\,s\langle\hat{\pi}^{ij}(0),\hat{\pi}^{ij}(s)\rangle}{\int_0^\infty\mathrm{d}s\langle\hat{\pi}^{ij}(0),\hat{\pi}^{ij}(s)\rangle},
 \\
 \label{eq:RT2-2}
 \tau_{J} &\equiv \frac{\int_0^\infty\mathrm{d}s\,s\langle\hat{J}^i(0),\hat{J}^i(s)\rangle}{\int_0^\infty\mathrm{d}s\langle\hat{J}^i(0),\hat{J}^i(s)\rangle},
\end{align}
which may give a clearer physical interpretation. 
Two remarks are in order here:
First, Eqs.~\eqref{eq:TC1-2} and \eqref{eq:TC2-2} are consistent with the Green-Kubo formula \cite{Jeon:1994if,Jeon:1995zm,Hidaka:2010gh}.
They actually take the same form as those of the Chapman-Enskog method.
Secondly, the forms of Eqs.~\eqref{eq:RT1-2} and \eqref{eq:RT2-2} give the time constant of the correlation function of the microscopic dissipative currents, which are physically natural forms as viscous relaxation times.


\begin{figure}[t]
 \centering
 \includegraphics[width=6.5cm]{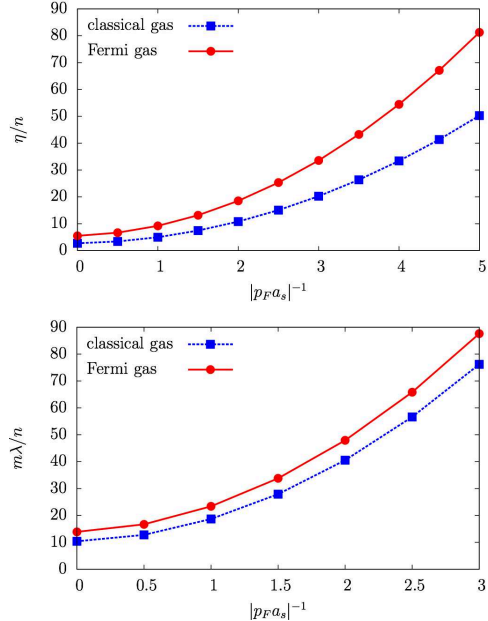}
 \caption{(Color online) Scattering length dependence of the shear viscosity (upper panel) and heat conductivity (lower panel) at the Fermi temperature. The blue square and red circle indicate the classical Boltzmann gas and Fermi gas, respectively.}
 \label{fig:a_1st_CF}
\end{figure}

\begin{figure}[t]
 \centering
 \includegraphics[width=6.5cm]{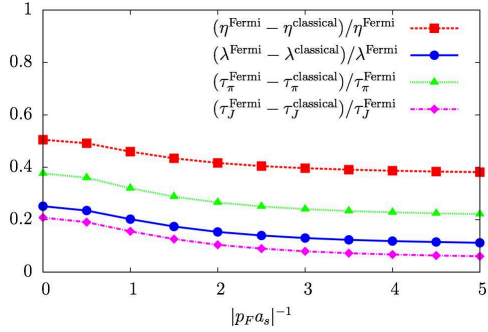}
 \caption{(Color online) Scattering length dependence of the quantum statistical effects for the shear viscosity (red square), heat conductivity (blue circle), the viscous relaxation times of the stress tensor (green triangle), and that of the heat conductivity (purple rhombus) are respectively shown at the Fermi temperature.}
 \label{fig:ratio_Q}
\end{figure}

\section{Quantum statistical effects}
\label{sec:sec4}
Aiming at the application to the ultracold Fermi gases realized in the cold-atom experiments, we examine the shear viscosity, heat conductivity, and the viscous relaxation times of the stress tensor and heat flow derived above.
We consider the s-wave scattering in the collision integral \eqref{eq:collision}.
Then, the scattering amplitude in Eq.~\eqref{eq:transition} is given by
\begin{align}
 \label{eq:cross_section}
 \mathcal{M} = \frac{4\pi}{a_s^{-1}+iq},
\end{align}
where $a_s$ is the s-wave scattering length and $\bs{q}=(\bs{p}-\bs{p}_1)/2$ is the incoming relative momentum.
In the present analyses based on Eq.~\eqref{eq:cross_section}, 
the transport coefficients become even functions of $a_s$, 
though they may loose such a symmetry
if the medium effects are taken into account \cite{Elliott:2014,Bluhm:2014uza}; 
the asymmetric behavior can be caused by the existence of a bosonic regime of preformed pairs for positive $a_s$, which is neglected in the present study and the incorporation of such an effect is beyond the scope of this work.
Thus the results shown below should be taken for the quantities rather for negative $a_s$.
We evaluate the expressions \eqref{eq:TC} and \eqref{eq:RT} with Eq.~\eqref{eq:cross_section} numerically in the following way:
Let $X_p \equiv \big[ L^{-1}\hat{\pi}^{ij}]_p$ or $\big[ L^{-1}\hat{J}^i \big]_p$,
which satisfy the linear equation $\big[ L X \big]_p = (\hat{\pi}^{ij},\hat{J}^i)_p$
where the explicit form of $L$ is known.
Then we discretize the last equation in the momentum space and convert it to a linear equation with a finite dimension,
which is readily solved numerically.
Then varying the dimension of the matrix, we check the convergence of the solution.
The details of the numerical procedure will 
be reported in Ref.~\cite{kikuchi:2015PTEP} (see \cite{de1980relativistic}, for the relativistic case). 
The exact evaluation of the transport coefficients makes it possible to examine the importance of the quantum statistical effects and the reliability of the RTA quantitatively as we discuss below.

\begin{figure}[t]
 \centering
 \includegraphics[width=6.5cm]{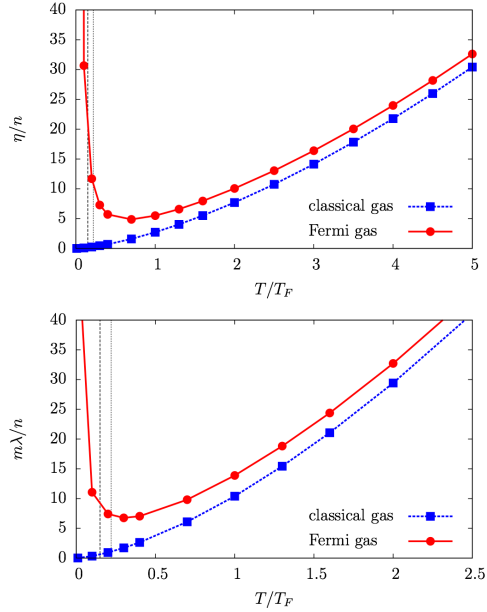}
 \caption{(Color online) Temperature dependence of the shear viscosity (upper panel) and heat conductivity (lower panel) at unitarity. The blue square and red circle indicate the transport coefficients of classical Boltzmann gas and the Fermi gas, respectively. The blue squares in the upper panel are on a curve $\eta/n = 2.77(T/T_F)^{3/2}$. The black dashed and dotted lines show the superfluid-transition temperature $T_c\simeq 0.167T_F$ \cite{Ku563} and the pairing-formation temperature $T^*\simeq 0.22T_F$ \cite{Wlazlowski:2013}, respectively.}
 \label{fig:T_1st}
\end{figure}

First, we calculate the shear viscosity and heat conductivity with varying the scattering length. The unitary limit is 
taken by $a_s\to \infty$, i.e., $(p_F a_s)^{-1}=0$, where $p_F$ is the Fermi momentum. For the classical Boltzmann gas, we set $a=0$ in Eq.~\eqref{eq:collision} and the corresponding equilibrium distribution function reads $f^\mathrm{eq}_p=\mathrm{e}^{-(|\bs{\delta p}|^2/2m-\mu)/T}$. For the Fermi gas, we take into account the quantum statistics by setting $a=-1$, and the equilibrium distribution function is $f^\mathrm{eq}_p=[\mathrm{e}^{(|\bs{\delta p}|^2/2m-\mu)/T}+1]^{-1}$.
The resultant data show that, as the scattering length increases, the viscous effects decrease (Fig.~\ref{fig:a_1st_CF}), and  the quantum statistical effects increase (Fig.~\ref{fig:ratio_Q}). The scattering-length dependence of the quantum statistical effects indicates that the quantum nature becomes apparent at unitarity, where the atomic gases are strongly correlated.
Figure~\ref{fig:T_1st} shows the temperature dependence of $\eta/n$ and $m\lambda/n$. 
The quantum statistical effect increases $\eta/n$ and $m\lambda/n$. The difference between 
the classical gas and the Fermi gas becomes larger as temperature decreases. 
In particular, at the temperature $T<T_F$, the differences become prominent and those of the Fermi gas diverge.
Two comments are in order here.
(i)~The quantum statistical effect is not negligible even above $T^*$, 
below which pairing effects dominate and our computation becomes invalid.
(ii)~The increases of the shear viscosity and heat conductivity for the Fermi gas 
at small temperature are naturally understood in terms of the Pauli blocking effect;
a good Fermi sphere is formed at low temperature and thus the elastic scattering rate other than the forward one
is so greatly suppressed due to the Pauli blocking that the energy 
and momentum transport become quite efficient, which implies the increase of the shear viscosity and heat conductivity.

\begin{figure}[t]
 \centering
 \includegraphics[width=6.5cm]{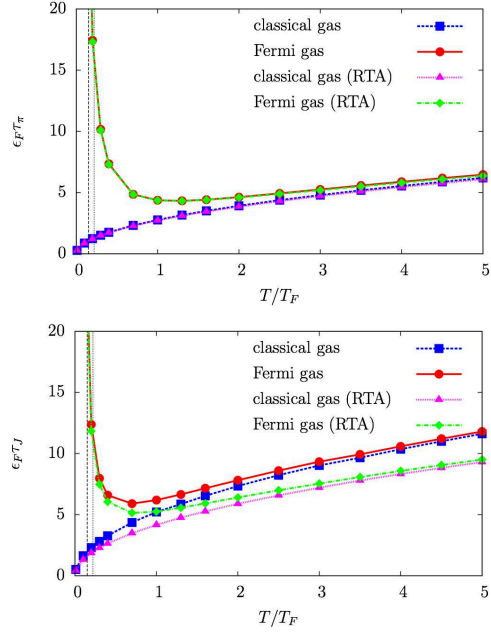}
 \caption{(Color online) Temperature dependence of the viscous relaxation times of the stress tensor (upper panel) and heat flow (lower panel) at unitarity. The blue square and red circle indicate the viscous relaxation times of classical Boltzmann gas and the Fermi gas which evaluated without the RTA, while the purple triangle and green rhombus indicate those evaluated with the RTA given by Eqs.~\eqref{eq:setting1} and \eqref{eq:setting2}. 
The black dashed and dotted lines show  $T_c$ and $T^*$, respectively.}
 \label{fig:T_tau}
\end{figure}

The viscous relaxation times exhibit similar behaviors as the transport coefficients. The quantum statistical effects increases at low temperature and near unitarity (see Figs.~\ref{fig:ratio_Q} and \ref{fig:T_tau}).

It seems that a misunderstanding prevails in the literature on the basis of the classical Boltzmann equation
that the shear viscosity calculated in the naive kinetic theory were to agree
with the experimental results at low temperature even close to $T^*$. 
Our results, however, have shown 
that the quantum statistical corrections are
huge and the resultant shear viscosity shows a rapid increase with lowering temperature 
which is obviously in contradiction to the experimental results. 
Thus we conclude that the naive kinetic approach is not 
valid for the description of the unitary Fermi gas at low temperature, 
where the system becomes strongly correlated, even above the phase transition temperature.

\begin{figure}[t]
 \centering
 \includegraphics[width=6.5cm]{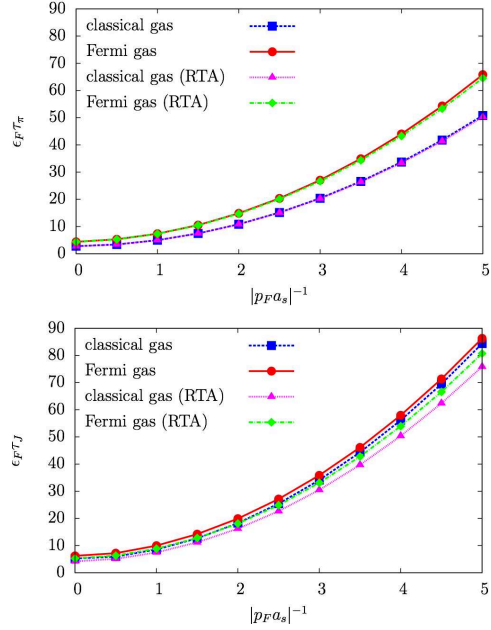}
 \caption{(Color online) Scattering length dependence of the viscous relaxation times of the stress tensor (upper panel) and heat flow (lower panel) at the Fermi temperature. The blue square and red circle indicate the viscous relaxation times of classical Boltzmann gas and the Fermi gas which evaluated without the RTA, while the purple triangle and green rhombus indicate those evaluated with the RTA given by Eqs.~\eqref{eq:setting1} and \eqref{eq:setting2}.}
 \label{fig:a_tau}
\end{figure}

\begin{figure}[htb]
 \centering
 \includegraphics[width=6.5cm]{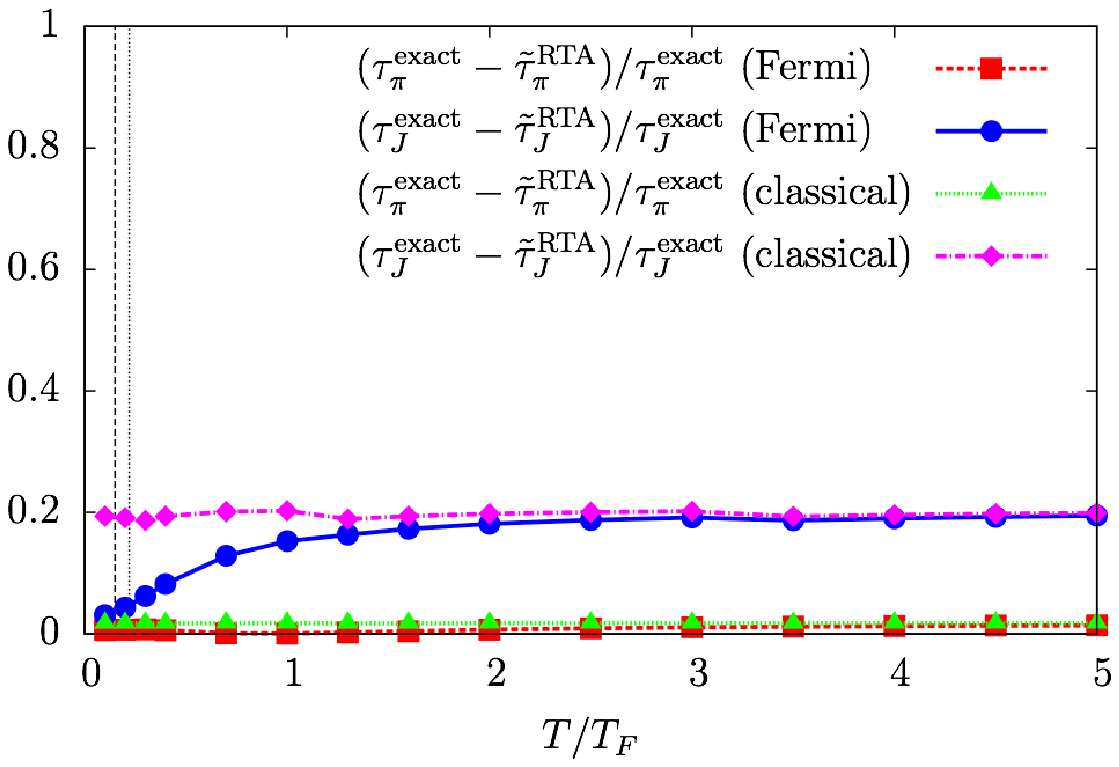}
 \caption{(Color online) Temperature dependence of the error caused by the RTA are shown for the shear viscosity (red square), heat conductivity (blue circle), the viscous relaxation times of the stress tensor (green triangle), and that of the heat conductivity (purple rhombus) are respectively shown at unitarity. 
The black dashed and dotted lines show $T_c$ and $T^*$, respectively.}
 \label{fig:ratio_RTA}
\end{figure}



\section{Precision test of RTA}
In the RTA, the collision integral in Eq.~\eqref{eq:collision} is replaced by
\begin{align}
 C[f]_p = -\frac{f_p-f^\mathrm{eq}_p}{\tau},
\end{align}
where $\tau$ is a free parameter which determines the time scale for a non-equilibrium system to relax toward the equilibrium states. 
Under the approximation, the shear viscosity and heat conductivity are calculated to be \cite{Bruun:2007,Braby:2011,Chao:2012}
\begin{align}
 \label{eq:transport_rel}
 \eta^{\mathrm{RTA}} = \tau P,
 \ \ \ 
 \lambda^{\mathrm{RTA}} = \frac{\tau}{12mT}\left(7Q-\frac{75P^2}{n}\right),
\end{align}
with the definition $Q \equiv \int_p \delta v^2 \delta p^2 f^\mathrm{eq}_p$.
In addition, it should be noted that the viscous relaxation times are given by $\tau$,
\begin{align}
 \label{eq:tau_rel}
 \tau^{\mathrm{RTA}}_\pi = \tau^{\mathrm{RTA}}_J = \tau.
\end{align}

However, the viscous relaxation times of the stress tensor and heat conductivity given by Eq.~\eqref{eq:RT} take the considerably different values (Fig.~\ref{fig:a_tau} and \ref{fig:T_tau}).
These results are interesting on its own, and also clearly contradict to Eq.~\eqref{eq:tau_rel} and indicate that the RTA should be modified so as to incorporate the multiple relaxation-time scales.
To this end, we determine the viscous relaxation times independently with the help of the relations Eq.~\eqref{eq:transport_rel} derived from the RTA and exact value of the shear viscosity and heat conductivity as follows:
We evaluate the viscous relaxation time of the stress tensor by
\begin{align}
 \label{eq:setting1}
 \tau_\pi = \frac{\eta^{\mathrm{exact}}}{P} \equiv \tilde{\tau}^\mathrm{RTA}_\pi.
\end{align}
While the viscous relaxation time of the heat conductivity is evaluated as
\begin{align}
 \label{eq:setting2}
 \tau_J = \frac{12mT\lambda^{\mathrm{exact}}}{(7Q-75P^2/n)} \equiv \tilde{\tau}^\mathrm{RTA}_J.
\end{align}
We note that $\eta^{\mathrm{exact}}$, $\lambda^{\mathrm{exact}}$, $\tau_\pi^{\mathrm{exact}}$, and $\tau_J^{\mathrm{exact}}$
denote the transport coefficients and viscous relaxation times which are respectively calculated from Eqs.~\eqref{eq:TC} and \eqref{eq:RT} in the exact manner in the present work.
In Figs.~\ref{fig:a_tau} and \ref{fig:T_tau}, we compare the viscous relaxation times with and without the RTA
and the numerical results of Eqs.~\eqref{eq:setting1} and \eqref{eq:setting2} actually behave similarly compared with the exact ones qualitatively.
It is remarkable that
$\eta^{\mathrm{exact}}/P$
well reproduces the viscous relaxation time of the stress tensor
$\tau^{\mathrm{exact}}_\pi$
for both the classical gas and the Fermi gas regardless of temperature and scattering length. On the other hand, the RTA still has the quantitative error and always underestimates the viscous relaxation time of the heat conductivity compared with those evaluated exactly.
It is noteworthy that the error of $\tau_J$ caused by the RTA for the Fermi gas decreases at low temperature in contrast to that for classical gas, which is independent of temperature (Fig.~\ref{fig:ratio_RTA}).
This behavior may be understood as follows:
The RTA is a kind of the linear approximation of the collision integral with respect to the deviation of the distribution function from the equilibrium one.
The deviation is attributed to the excitations of quasi-particles due to the nonequilibrium process and decreases in the cold fermionic gases 
since the Pauli-blocking effect suppresses the excitation of the quasi-particles inside the Fermi sphere.
Therefore the linear approximation works well and the RTA reproduces the exact value of the viscous relaxation time of the heat conductivity.


\section{Conclusion}
\label{sec:sec5}
We have calculated in a full numerical way the shear viscosity, heat conductivity and viscous relaxation times of the stress tensor and heat flow using the microscopic expressions which are derived by the RG method.
It is notable that the shear viscosity and heat conductivity by the RG method have the same expression as those by the Chapman-Enskog method, and the viscous relaxation times take new but natural forms.
Any approximation has not been used in the numerical evaluation of these quantities in the present work, which has given the exact values based on the Boltzmann equation.
We have found that the Fermi statistics makes significant contributions to the transport coefficients and viscous relaxation times at low temperature or large scattering length.
Furthermore, by using the numerical results we have for the first time examined the reliability of the relations $\tau_\pi=\eta/P$ and $\tau_J = 12mT\lambda/(7Q-75P^2/n)$, which are derived with recourse to the RTA and used rather extensively. The resulting data have shown that the ratio of the shear viscosity and pressure well agrees with the viscous relaxation time. This agreement is consistent with Ref.~\cite{Schafer:2014} and suggests the reliability of the RG method. Furthermore, the results encourage us to use the relation $\tau_\pi=\eta/P$, which greatly simplifies the evaluation of the viscous relaxation time of the stress tensor. On the other hand, the latter relation for the viscous relaxation time does not appear to be reliable. Thus, we should use the value evaluated by Eq.~\eqref{eq:RT} instead of Eq.~\eqref{eq:transport_rel} for the viscous relaxation time of the heat conductivity.
The detailed analyses of this work will be reported in Ref.~\cite{kikuchi:2015PTEP}.
Investigation of the time-evolution of fluids with these transport coefficients is needed to examine quantitative significance of the difference of $\tau_J$ evaluated with and without the RTA, and it is left as a future project.


\section*{Acknowledgments}
Y.K. is supported by the Grants-in-Aid for JSPS fellows (No.15J01626). 
T.K. was partially supported by a
Grant-in-Aid for Scientific Research from the Ministry of Education,
Culture, Sports, Science and Technology (MEXT) of Japan
(Nos. 20540265 and 23340067),
by the Yukawa International Program for Quark-Hadron Sciences.


\end{document}